\documentclass[conference]{IEEEtran}
\usepackage{verbatim}  
\usepackage{epsf}
\usepackage{epsfig}
\usepackage{epstopdf}
\usepackage{enumerate}
\usepackage{times}
\usepackage{amsmath}
\usepackage{amssymb}
\usepackage{cite}

\IEEEoverridecommandlockouts

\usepackage{algorithm}
\usepackage{algpseudocode}

\usepackage[hidelinks,bookmarks=false]{hyperref}

\title{The Effect of Channel Uncertainty on Max-Min Goodput}

\author{\IEEEauthorblockN{Mostafa Medra$^\ast$ \quad Andrew W. Eckford$^\dagger$ \quad Raviraj Adve$^\ast$}
\IEEEauthorblockA{$^\ast$ Department of Electrical and Computer Engineering, University of Toronto, ON, Canada. \\
$^\dagger$ Department of Electrical Engineering and Computer Science, York University, Toronto, ON, Canada.}
\thanks{This research is supported by TELUS Canada and the Natural Sciences and Engineering Research Council (NSERC) of Canada.}
}

\begin{document}
\maketitle

\begin{abstract}

In this paper, we consider the effect of channel uncertainty on the rates reliably delivered to the users; i.e., the goodput. After the base station (BS) designs a set of beamformers for a specific objective, the BS must select the operating or transmission data rate for each user. However, under channel uncertainty, higher transmission rates cause higher outage probability, and the delivered rate drops. Since lower rates are not desirable, one must balance between  the transmission rate and outage. In this paper, we first explain how approximating the PDF of a quadratic form with a positive definite matrix can be used to obtain the outage probability for any set of beamfomers and transmission rate. Then we focus on the specific case of maximizing the minimum delivered rate, where we modify a robust beamforming approach to maximize the resulting goodput. We then derive iterative closed-form expressions for this case. The simulation results illustrate the efficacy of our analysis and the significant gains that can be obtained by optimizing the goodput metric.

\end{abstract}

\section{Introduction}

The presence of channel state information at the transmitter (CSIT) plays an important role in enhancing the performance of multiple-input multiple-output (MIMO) systems \cite{CapacityofMultiantennaGaussian}.  However, in practice, the channels between the users and the base station (BS) must be estimated, often using training. For example, in time division duplexing (TDD) systems that estimation is typically performed during the uplink training phase. Another example is the frequency division duplexing (FDD) systems where each receiver estimates its channel and feeds back a quantized version of that estimate to the BS. Crucially, for our purposes, regardless of the system used, the available CSIT is noisy.

When considering very long time intervals, the ergodic capacity can be used to describe the system performance. However, for delay-limited applications the outage capacity is a preferred metric~\cite{Efficientuseoffadingcorrelations,simon2005digital}. In this case, we investigate each data block separately. If it is successfully decoded, we proceed to the next data block. Otherwise, we have an outage and the packet must be retransmitted. That can be achieved using automatic-repeat-request algorithms; e.g.,~\cite{AnARQSchemewithMemory,Automaticrepeatrequest}, where the user sends an ``acknowledged" (ACK) when the packet is successfully decoded or ``not acknowledged" (NAK) when it requires retransmission. These ARQ protocols, therefore, provide reliability to the transmission.

When we define the outage capacity as the maximum transmission rate that can guarantee a certain outage probability, that definition does not provide us with the best transmission rate nor outage. The transmission rate that is set at the transmitter determines the outage probability, and, in turn, describes how much of that rate can be realized at the receiver. When we have a reliable system; e.g., ARQ-based systems, the throughput is named goodput~\cite{ThroughputOptimalPrecoding}. By measuring the rate actually realized at the receiver, goodput provides an effective metric to measure system performance~\cite{Rateadaptationvialink,RobustRatePowerandPrecoder}.

Using goodput as the performance metric raises a distinction between the transmission rate and the received rate. A high transmission rate results in a high outage probability (due to channel uncertainty; i.e., error in channel estimates) and most of the rate is lost in outage. On the other hand if the transmission rate is set too low, the received rate will be low. In~\cite{TheEffectofChannelEstimation}, the authors  first select the user with the best channel, then for that user, choose a back off factor, from the maximum achievable rate, between zero and one. The factor is chosen to to maximize the goodput; this is, in fact, a common approach in most works. In~\cite{GoodputOptimalRateAdaptation}, the authors define a, similar, utilization factor that is optimized for a single-input single-output (SISO) system. The authors in~\cite{ThroughputOptimalPrecoding} provide goodput maximization for a single-user MISO system. Optimization for the point-to-point MIMO goodput was presented in~\cite{RobustRatePowerandPrecoder}. In \cite{OnimperfectCSIforthedownlink}, the rate back off is optimized for the downlink of a two-tier network. In \cite{JointRateandPowerAllocation}, the  throughput maximization was done for a system using non-orthogonal multiple access (NOMA).  In \cite{DesignandanalysisofmultiuserSDMA}, the authors focused on goodput maximization for multi-user MISO system. Due to the mathematically challenging problem, orthogonal precoding is used with equal power loading and the codebook used is orthonormal. In \cite{Rateadaptationvialink}, the authors used the ACK and NAK signals available from the ARQ protocol to adapt the transmission rate in order to maximize the long-term expected goodput.

This paper considers the goodput optimization problem, under channel uncertainty, in the context of a \emph{multiuser} MISO system. The BS uses zero-forcing (ZF) beamforming to service multiple users. Specifically, we optimize max-min rate delivered to users with a certain guarantee. Further, we consider the more general case of hybrid-ARQ (HARQ) protocols~\cite{Forwarderrorcorrection}, wherein packets in outage are not dropped and retransmitted, but augmented with parity bits. This allows the receiver to recover some part of the data rate associated with the packets in outage.

In analyzing goodput, calculating the outage probability requires the probability density function (PDF) of a quadratic term with indefinite matrix. The PDF of a quadratic term with positive definite matrix has been developed in~\cite{Ontheinversionofcertain}. Then, the quadratic term with indefinite matrix was decomposed to two independent quadratic forms with definite matrices in~\cite{Anaccurateapproximationtothedistribution}. These mathematical tools provide an approach to obtain the outage probability for any beamforming vectors and transmission rates.

The key contribution in this paper is to underline the significant gains to be had in \emph{delivered} rates to users using the goodput framework. We show that the goodput is highly dependent on the transmission rate, and significant gains can be obtained by properly choosing that rate. We then show how a robust approach can be modified such that we can balance between the transmission rate and outage. We focus on the case of maximizing the minimum delivered rate, and provide iterative closed-form expressions for this case.

\section{System Model}

We consider a narrowband  multiuser MISO downlink system with an $N_t$-antenna BS and $K$ single-antenna users. The transmitted signal  $\mathbf{x}_t$ is designed using linear beamforming such that $\mathbf{x}_t= \sum_{k=1}^K\mathbf{w}_k s_k$, where $s_k$ is the data symbol for user $k$, and $\mathbf{w}_k$ the associated beamformer. The received signal at user $k$ is
\begin{equation}\label{rcvd_sig}
    y_k= \mathbf{h}_k^H \mathbf{w}_k s_k + \textstyle\sum_{j \neq k}\mathbf{h}_k^H \mathbf{w}_j s_j + n_k,
\end{equation}
where $\mathbf{h}_k^H$ denotes the actual channel between the BS and receiver $k$, and $n_k$ represents the additive zero-mean circular complex Gaussian noise at that user. The signal-to-interference-and-noise ratio (SINR) value at user $k$ is
\begin{equation}\
    \text{SINR}_k =  \frac{\mathbf{h}_k^H \mathbf{w}_k \mathbf{w}_k^H \mathbf{h}_k}{\mathbf{h}_k^H  (\sum_{j \neq k}\mathbf{w}_j \mathbf{w}_j^H) \mathbf{h}_k + \sigma_k^2} , \\
\end{equation}
where $\sigma_k^2$ is the noise variance at receiver $k$. In the case of zero-mean independent data symbols of normalized power, the average transmission power is $\textstyle\sum_{k=1}^K  \mathbf{w}_k^H \mathbf{w}_k$. In this paper, we will assume that we have a total power constraint; $\textstyle\sum_{k=1}^K  \mathbf{w}_k^H \mathbf{w}_k \leq P_t$, where $ P_t$ is the maximum power allowed for transmission.

The SINR at the user is important in determining the rate that can be transmitted with tolerable bit-error rate. However, that value depends on the channels being known at the BS. In reality, the channels are estimated and channel uncertainty is inevitable. In this paper, we will consider systems with additive channel uncertainty modeled as
\begin{equation}\label{uncertainty}
   \mathbf{h}_k= \mathbf{h}_{e_k} +\mathbf{e}_k,
\end{equation}
where $\mathbf{h}_{e_k}$ is the BS's estimate of the channel to user $k$, and $\mathbf{e}_k \backsim \mathcal{CN} (\boldsymbol\mu_k,\mathbf{C}_k)$, denoting $\mathbf{e}_k$ to be complex Gaussian with mean $\boldsymbol{\mu}_k$ and covariance $\mathbf{C}_k$.

Based on this channel uncertainty, the outage probability $\delta_k= \mathbb{P}[\text{SINR}_k \leq \gamma_k$], where $\gamma_k$ is the SINR target that sets the transmission rate and $\mathbb{P}[\cdot]$ denotes probability. The outage probability is equivalent to $\mathbb{P}[\mathbf{h}_k^H \mathbf{Q}_k \mathbf{h}_k - \sigma_k^2 \leq 0]$, where
\begin{equation}
\begin{aligned}
    \mathbf{Q}_k &= \mathbf{w}_k \mathbf{w}_k^H/\gamma_k-\textstyle\sum_{j \neq k} \mathbf{w}_j \mathbf{w}_j^H \\
                 &= \beta_k  \mathbf{u}_k \mathbf{u}_k^H/\gamma_k-\textstyle\sum_{j \neq k} \beta_j \mathbf{u}_j \mathbf{u}_j^H,
     \end{aligned}
\end{equation}
with $\beta_k=\| \mathbf{w}_k \|^2$, and $ \mathbf{u}_k$ the unit norm vector in the direction of $ \mathbf{w}_k$.
Using the additive channel uncertainty model in \eqref{uncertainty}, we have that the outage probability, $\delta_k$, is $\mathbb{P}[f_k(\mathbf{e}_k) \leq 0]$ with
\begin{equation}\label{SINR_reformulation}
   f_k(\mathbf{e}_k)=\mathbf{h}_{e_k}^H \mathbf{Q}_k \mathbf{h}_{e_k} + 2 \mathcal{R}\left(\mathbf{e}_k^H  \mathbf{Q}_k \mathbf{h}_{e_k}  \right) + \mathbf{e}_k^H  \mathbf{Q}_k \mathbf{e}_k  - \sigma_k^2,
\end{equation}
where $\mathcal{R}\left(\cdot\right)$ denotes the real part of a complex number.

Outage constraints are generally intractable even when the uncertainty is Gaussian~\cite{Optimalpowercontrol,Probabilisticallyconstrained,OutageConstrained}. We will show in this paper how we can calculate such a probability and how to use that value to enhance the system performance.

\subsection{Goodput Rate: The Objective Function}

If the BS ignored the channel uncertainty, it would transmit at a rate $\log(1+\gamma_k)$ supported by the SINR value $\gamma_k$. Typically, the beamforming vectors reduce the interference terms in the denominator of the SINR so that we can have  high SINR values. When we have channel uncertainties, extra interference terms in the form of $\mathbf{e}_k^H  (\sum_{j \neq k}\mathbf{w}_j \mathbf{w}_j^H) \mathbf{e}_k$ will be added to the denominator which will decrease the SINR value and the user will suffer from high outage.

If we denote  the rate transmitted to user $k$ as $R_k$, this rate is the goodput conditioned on there being no outage; this happens with probability $1 - \delta_k$. However, when an outage happens, the user will ask for the packets to be retransmitted again. If the packet during outage is just discarded, then the rate retrieved from that packet is $\eta=0$. When advanced HARQ algorithms that make use of the previous sent packets are implemented, some of the transmitted  rate can be retrieved, $\eta R_k$, where $\eta < 1$. Accordingly, the sum goodput can be written as
\begin{equation}\label{goodput_obj}
R^{(g)} =  \sum_{k=1}^K (1-\delta_k) R_k +  \eta \delta_k R_k.
\end{equation}
Since we focus here on the case of max-min, the objective function simplifies to $R^{(g)} =  \sum_{k=1}^K (1-\delta_k) R +  \eta \delta_k R$.

The differences between this objective and the sum-rate (SR) approach are important to  highlight. To maximize SR, we maximize  $ \log(1+\gamma_k)$; however, for goodput, a higher $R_k$ results in a higher outage, $\delta_k$; i.e., there is a tradeoff between transmitted rate and outage. Furthermore, we note that the value of $\eta$, related to the HARQ, sets the importance of preventing the occurrence of an outage (for simplicity, in this paper, we assume that $\eta$ is constant).

\section{Effect of Channel Uncertainties on Outage}

Once the BS chooses a transmission rate, $R_k$, to user $k$, the outage probability, $\delta_k$, is given by $\mathbb{P}[f_k(\mathbf{e}_k)<0]$.  We now show how to efficiently approximate this expression using the approximate PDF of positive definite quadratic forms~\cite{Anaccurateapproximationtothedistribution,Ontheinversionofcertain}.

Since $\mathbf{e}_k \backsim \mathcal{CN} (\boldsymbol\mu_k,\mathbf{C}_k)$, we can write $\mathbf{e}_k=\boldsymbol\mu_k+\mathbf{C}^{1/2}_k \mathbf{z}_k$, where  $\mathbf{z}_k \backsim \mathcal{CN} (\boldsymbol 0,\mathbf{I})$.
Accordingly, we can rewrite $f_k(\mathbf{e}_k)$ as

\begin{subequations}\label{matrix_decomp}
\begin{align}
f_k(\mathbf{e}_k) &=\mathbf{h}_k^H \mathbf{Q}_k \mathbf{h}_k - \sigma_k^2 \\
&= (\mathbf{h}_{e_k}+\mathbf{e}_k)^H \mathbf{Q}_k (\mathbf{h}_{e_k}+\mathbf{e}_k)- \sigma_k^2 \\
&= (\mathbf{h}_{e_k}+\boldsymbol\mu_k+\mathbf{C}^{1/2}_k \mathbf{z}_k)^H \mathbf{Q}_k (\mathbf{h}_{e_k}+\boldsymbol\mu_k+\mathbf{C}^{1/2}_k \mathbf{z}_k) \nonumber \\ & \hspace*{1.5in} -\sigma_k^2 \\
&= (\mathbf{C}^{-1/2}_k(\mathbf{h}_{e_k}+\boldsymbol\mu_k)+ \mathbf{z}_k)^H \mathbf{C}^{1/2}_k \mathbf{Q}_k \mathbf{C}^{1/2}_k  \nonumber \\                 & \hspace*{0.7in} \times(\mathbf{C}^{-1/2}_k(\mathbf{h}_{e_k}+\boldsymbol\mu_k)+ \mathbf{z}_k)- \sigma_k^2 \\
&= (\mathbf{P}^H_k \mathbf{C}_k^{-1/2}(\mathbf{h}_{e_k}+\boldsymbol\mu_k)+ \mathbf{P}^H_k \mathbf{z}_k)^H  \boldsymbol\Delta_k \nonumber \\  \label{SINR_sep}
& \hspace*{0.2in} \times(\mathbf{P}^H_k \mathbf{C}^{-1/2}_k (\mathbf{h}_{e_k}+\boldsymbol\mu_k)+ \mathbf{P}^H_k \mathbf{z}_k)- \sigma_k^2,
 \end{align}
\end{subequations}
where $\mathbf{C}^{1/2}_k \mathbf{Q}_k \mathbf{C}^{1/2}_k = \mathbf{P}_k \boldsymbol\Delta_k \mathbf{P}^H_k,$  $\boldsymbol\Delta_k$ is a diagonal matrix, and $\mathbf{P}_k$ is an orthonormal matrix. Now in~\eqref{SINR_sep} we observe that the term $\mathbf{P}^H_k \mathbf{C}^{-1/2}_k (\mathbf{h}_{e_k}+\boldsymbol\mu_k)$ is constant, and that, since $\mathbf{P}_k$ is an orthonormal matrix, $\mathbf{P}^H_k \mathbf{z}_k$ is a vector of independent normal random variables. Accordingly, the quadratic form in \eqref{SINR_sep} can be divided into the difference of two independent quadratic positive definite forms; $(\mathbf{b}_1+\mathbf{z}_1)^H\boldsymbol\Delta_1 (\mathbf{b}_1+\mathbf{z}_1) - (\mathbf{b}_2+\mathbf{z}_2)^H\boldsymbol\Delta_2 (\mathbf{b}_2+\mathbf{z}_2)-\sigma_k^2$. The first positive definite term corresponds to the positive values in $\boldsymbol\Delta$, while the second corresponds to the negative values. Since the individual terms are independent, the overall PDF can be obtained directly by using convolution.

\subsection{Approximating the PDF of a Positive Definite Quadratic Form}\label{PDF_PDM}

In this section, we will summarize the steps required to approximate the PDF of a  positive definite quadratic form $\mathbf{x}^T \mathbf{A} \mathbf{x}$ as shown in \cite{Anaccurateapproximationtothedistribution}. The approximation is based on obtaining the moments of the quadratic form from its cumulants by using a recursive formula, then using those moments to approximate the density function of the positive definite quadratic form. That approximation uses Laguerre polynomials and makes use of their orthogonality to obtain the polynomials' weights. Those weights are selected such that the first $d$ moments of the original and approximated PDF are the same, where $d$ sets the degree, and, accordingly, accuracy of approximation.

For  $\mathbf{x} \backsim \mathcal{N} (\boldsymbol\mu_x,\mathbf{C})$ and a real positive definite matrix $\mathbf{A}$, the $i$th cumulant, $c_i$, of $\mathbf{x}^T \mathbf{A} \mathbf{x}$  is \cite{Quadraticformsinrandom}
\begin{equation}\label{cumulants}
   c_i=2^{i-1} i! (\text{tr}(\mathbf{A}\mathbf{C})^i/i+ \boldsymbol\mu_x^T (\mathbf{A}\mathbf{C} )^{i-1} \mathbf{A}\boldsymbol\mu_x),
\end{equation}
where $\mathrm{tr}(\cdot)$ denotes the trace function.
Using the calculated cumulants, we can obtain the $i$th moment $\boldsymbol\chi_i$ as~\cite{Arecursiveformulationoftheoldproblem}
\begin{equation}\label{moments}
   \chi_i= \sum_{k=0}^{i-1} \frac{(i-1)!}{(i-k-1)!k!} c_{i-k} \chi_k.
\end{equation}
We define $\beta_x =\chi_2/\chi_1 - \chi_1, \nu=\chi^2_1/(\chi_2 - \chi_1^2) -1, d_{i,k}= \frac{ (-1)^k \Gamma(i+\nu+1)  }{ (i-k)! k! \Gamma(\nu+k+1)}, \eta_i=\frac{\Gamma(\nu+1) i!}{\Gamma(\nu+i+1)} \sum_{k=0}^i d_{i,k} \chi_k/\beta^k_x,$ and $\xi_k=\sum_{i=k}^d  \eta_i d_{i,k}.$
Using those definitions, the approximated density at $y$ can be written as \cite{Ontheinversionofcertain}
\begin{equation}\label{pdf_approx}
   f_a(y)=\frac{y^\nu e^{-y/\beta_x}}{\beta_x^{\nu} \Gamma(\nu+1)} \sum_{k=0}^d \xi_k y^k/\beta^{k+1}_x.
\end{equation}

\subsection{Approximating the PDF of $f_k(\mathbf{e}_k)$}

To approximate the PDF of $f_k(\mathbf{e}_k)$, the only remaining point to deal with is the fact that the quadratic term $\mathbf{h}_k^H \mathbf{Q}_k \mathbf{h}_k$ consists of complex vector $\mathbf{h}_k$ and matrix $\mathbf{Q}_k $, while the derivations of the approximate PDF for the form $\mathbf{x}^T \mathbf{A} \mathbf{x}$ assumed real values only. However, a complex quadratic form can be directly decoupled to a real quadratic form of twice the dimensions as follows
\begin{equation}\label{real_imag_decop}
   \mathbf{h}_k^H \mathbf{Q}_k \mathbf{h}_k=
\left[\begin{array}{l}\mathcal{R}(\mathbf{h}_k)\\\mathcal{I}(\mathbf{h}_k)\end{array}\right]^T
\left[\begin{array}{c c}\mathcal{R}(\mathbf{Q}_k)&-\mathcal{I}(\mathbf{Q}_k)\\ \mathcal{I}(\mathbf{Q}_k )&\mathcal{R}(\mathbf{Q}_k )\end{array}\right]
\left[\begin{array}{l}\mathcal{R}(\mathbf{h}_k)\\\mathcal{I}(\mathbf{h}_k)\end{array}\right],
\end{equation}
where $\mathcal{I}(\cdot)$ denotes the imaginary part of a complex number.

Now, the PDF of $f_k(\mathbf{e}_k)$ can be efficiently approximated by obtaining the PDFs of its underlying definite quadratic components. The steps for doing so are summarized in Alg.~\ref{Alg1}.

\begin{algorithm}
\caption{Approximating the PDF of $f_k(\mathbf{e}_k)$}
\label{Alg1}
\begin{algorithmic}[1]
\State Decompose $f_k(\mathbf{e}_k)$ into two quadratic forms $(\mathbf{b}_i+\mathbf{z}_i)^H\boldsymbol\Delta_i (\mathbf{b}_i+\mathbf{z}_i)$ using \eqref{matrix_decomp}, $i=1,2$.
\State Do steps 3-5 for each of the quadratic forms.
\State  Use \eqref{real_imag_decop} to obtain the real representation $\mathbf{x}^T \mathbf{A} \mathbf{x}$ for $(\mathbf{b}_i+\mathbf{z}_i)^H\boldsymbol\Delta_i (\mathbf{b}_i+\mathbf{z}_i)$.
\State Compute the moments $\chi_i$ using \eqref{cumulants}, \eqref{moments}.
  \State Calculate the constants and  weights to obtain $f_a^i(\cdot)$ using \eqref{pdf_approx}.
  \State Obtain the approximate PDF $f_a(\cdot)$ as the convolution of $f_a^1(\cdot)$ and $f_a^2(\cdot)$.
  \State The outage probability is the probability that  $f_a(\cdot) < \sigma_k^2.$
\end{algorithmic}
\end{algorithm}

The technique described in Alg.~\ref{Alg1} provides the outage probability for a given transmission rate for any set of beamforming vectors. That enables us to find the different outage probabilities for a range of transmission rates, and obtain the goodput $R^{(g)}$ in~\eqref{goodput_obj} for those rates.

\section{Robust Beamforming to Optimize Max-Min Goodput}

The goodput objective in \eqref{goodput_obj} combines the hardness of rate algorithms in addition to the intractability of the outage probability constraints. However, without both, we are not measuring the actual rate at the receiver side. To untangle this problem, we will consider a two-layer approach. For the outer layer, we will modify a heuristic approach to control the outage, then use insights from a special case to approximate the outage probability $\delta$. Then for the inner layer, we will use that $\delta$ to obtain closed-form expressions for the max-min rates.

\subsection{The Heuristic Approach and the Proposed Modification}

When dealing with channel uncertainty, a well-known heuristic is to combat the expected resulting interference by adding additional noise terms for user $k$, specifically $\mathbb{E}[\mathbf{e}_k^H  (\sum_{j \neq k}\mathbf{w}_j \mathbf{w}_j^H) \mathbf{e}_k]$, where $\mathbb{E}[\cdot]$ denotes expectation. This additional noise term can be written as
\begin{subequations}
\begin{align}
n_{e_k}^2&= \mathbb{E}[\mathbf{e}_k^H  (\sum_{j \neq k}\mathbf{w}_j \mathbf{w}_j^H) \mathbf{e}_k] \\
&= \mathbb{E}[(\boldsymbol\mu_k+\mathbf{C}_k^{1/2} \mathbf{z}_k)^H (\sum_{j \neq k}\mathbf{w}_j \mathbf{w}_j^H) (\boldsymbol\mu_k+\mathbf{C}_k^{1/2} \mathbf{z}_k))] \\
\pagebreak
&=\boldsymbol\mu_k^H (\sum_{j \neq k}\mathbf{w}_j \mathbf{w}_j^H)\boldsymbol\mu_k+  \mathbb{E} [\sum_{j \neq k} \mathbf{w}_j^H \mathbf{C}_k^{1/2} \mathbf{z}_k \mathbf{z}_k^H \mathbf{C}_k^{1/2}\mathbf{w}_j] \\
&=\boldsymbol\mu_k^H (\sum_{j \neq k}\mathbf{w}_j \mathbf{w}_j^H)\boldsymbol\mu_k+ \sum_{j \neq k} \mathbf{w}_j^H \mathbf{C}_k \mathbf{w}_j \\
&=\boldsymbol\mu_k^H (\sum_{j \neq k} \beta_j \mathbf{u}_j \mathbf{u}_j^H)\boldsymbol\mu_k+ \sum_{j \neq k} \beta_j \mathbf{u}_j^H \mathbf{C}_k \mathbf{u}_j.
 \end{align}
\end{subequations}
For a given set of beamforming vectors, $n_{e_k}^2$ is linear in each $\beta_k$. Further, when $\mathbf{e}_k \backsim \mathcal{CN} (\boldsymbol 0, \sigma_{e_k}^2 \mathbf{I})$, we have the simplified factor $n_{e_k}^2=\sum_{j \neq k} \beta_j  \sigma_{e_k}^2.$

Adding this term to the noise essentially assumes that the additional interference is Gaussian (the worst-case from an information theoretic perspective) and infinite block lengths. However, this approach must be visited in the framework of outage probability.

We observe that, with channel uncertainty, we have two factors affecting the SINR. The first is that the true signal power is $\mathbf{h}_k^H \mathbf{w}_k \mathbf{w}_k^H \mathbf{h}_k$, while we only have an estimated channel $\mathbf{h}_{e_k}^H$ instead. However, this effect is small as the magnitude of $\mathbf{h}_{e_k}^H$ should be significantly larger than the channel estimation error $\mathbf{e}_k$ for proper operation \cite{MIMObroadcast}. The second effect is more important which is the effect of the channel estimation error on the interference terms. When assuming ZF beamforming,  the term $\mathbf{e}_k^H  (\sum_{j \neq k}\mathbf{w}_j \mathbf{w}_j^H) \mathbf{e}_k$ can significantly degrades the overall SINR. This term is what the heuristic use to combat the noise, by adding its expected value to the noise level.

Note that, for fixed beamformers, we can obtain the PDF of the term $\mathbf{e}_k^H  (\sum_{j \neq k}\mathbf{w}_j \mathbf{w}_j^H) \mathbf{e}_k$ as shown in Section~\ref{PDF_PDM} since it is a quadratic form with a positive definite matrix, and that PDF could be used to approximate the outage probability for a certain SINR target. The advantage of this technique is that we have only one PDF to estimate, as that term is not a  function of the SINR value. However, when the noise level is very high, the effect of interference terms on outage decreases and the effect of the channel error in the SINR numerator becomes more dominant.

\subsection{Insights from a Special Case}

In the special case of only one significant interferer indexed $j$ and white estimation error, the PDF of $\mathbf{e}_k^H \mathbf{w}_j \mathbf{w}_j^H \mathbf{e}_k$, where  $\mathbf{e}_k \backsim \mathcal{CN} (\boldsymbol 0, \sigma_{e_k}^2 \mathbf{I})$, can be obtained using  \eqref{cumulants}, \eqref{moments} and \eqref{pdf_approx} as follows:
\begin{equation}\label{err_pdf_approx}
   f_a(y)= e^{-y/\beta_x} /\beta_x,
\end{equation}
where $\beta_x= \beta_j  \sigma_{e_k}^2$. In this case, $\mathbb{P}[Y > a\beta_x] = e^{-a}$ where $a$ is a threshold factor. Accordingly, we can choose the value of $a$ to balance between maximizing the rate and minimizing the outage to maximize the overall goodput. Note that in the case of one strong interferer, $a \beta_x $ is approximately $a$ times $n_{e_k}^2.$ This inspires us to modify the heuristic approach such that it provides robustness by \emph{scaling the average interference power} $n_{e_k}^2$ by a factor of $a$ to combat interference. This way, we control the outage probability by choosing $a$, and we approximate the outage probability  by $e^{-a}$.

Using our proposed two-layer approach, for a given value of $a$,  the inner layer can be written generally in the SR case as follows
\begin{subequations}\label{heurisric}
\begin{align}
      \max_{\substack{\gamma_k, \mathbf{w}_k}} \quad  & \sum_{k=1}^K  \log(1+\gamma_k) \\
      \text{s.t.} \quad &  \mathbf{h}_{e_k}^H \mathbf{Q}_k \mathbf{h}_{e_k} - a n_{e_k}^2  - \sigma_k^2 \geq 0,    \\
      & \sum_k  \beta_k  \leq P_t.
\end{align}
\end{subequations}
The problem in \eqref{heurisric} is hard to solve even when $a n_{e_k}^2$ is set to zero. A typical approach for SR algorithms is to use ZF directions, which allows for an optimal water-filling algorithm for the power loading. However, when the extra terms $a n_{e_k}^2$ are added, the problem is non-convex, even when ZF directions are assumed. On the other hand, when we investigate the case of maximizing the minimum rate when ZF beamforming is used, the problem in \eqref{heurisric} can be optimally solved. Since all the constraints are linear in $\beta_k$, we can use a  bisection search for the common SINR $\gamma$ and in each iteration the linear program is checked for feasibility. If the problem is to be solved for a general SR objective, one possible way is to use the expected value of $\beta_k=P_t/K$  in $n_{e_k}$ to ensure that $a n_{e_k}^2$ is constant. In this case, water-filling can be optimally used to obtain the power loading.

\subsection{Max-Min Closed-From Expressions}\label{sect_heur_maxmin}

We stated that the problem in \eqref{heurisric} can be solved by bisection search if the objective is maximizing the minimum rate. Here, we provide  efficient closed-form expressions to do so when ZF directions are used.

We observe that we can write the set of $K$ constraints $ \mathbf{h}_{e_k}^H \mathbf{Q}_k \mathbf{h}_{e_k} - a n_{e_k}^2  - \sigma_k^2 \geq 0$ in a matrix form as follows
\begin{equation}\label{A_eqn}
   \mathbf{A} \boldsymbol{\beta} \geq \boldsymbol{\sigma}^2,
\end{equation}
where $\boldsymbol{\beta}=[\beta_1, \beta_2,..., \beta_K]^T$, $\boldsymbol{\sigma}=[\sigma_{1}, \sigma_{2},..., \sigma_{K}]^T$, $\mathbf{[A]}_{ii}= | \mathbf{h}_{e_i}^H {\mathbf{u}}_i |^2/\gamma_i$, and $\mathbf{[A]}_{ij}= - a \sigma_{e_i}^2$,  $\forall i \neq j$. The power loading satisfying this constraint by equality is $  \boldsymbol{\beta} = \mathbf{A}^{-1} \boldsymbol{\sigma}^2.$
If we let $\boldsymbol{1}$ denote a column vector of ones, then the power constraint can be written as
$$  P_t \geq \boldsymbol{1}^H \mathbf{A}^{-1} \boldsymbol{\sigma}^2.$$
We observe that the matrix $\mathbf{A}$ can be written as a  diagonal matrix $\mathbf{B}$ that is perturbed by a rank one update $\mathbf{s} \boldsymbol{1}^H$, where $\mathbf{s}_k=- a \sigma_{e_k}^2$. Accordingly, we have \cite{sherman1949adjustment}
\begin{subequations}
\begin{align}
\boldsymbol{1}^H \mathbf{A}^{-1} \boldsymbol{\sigma}^2 & = \boldsymbol{1}^H \Bigl(\mathbf{B}^{-1} + \frac{\mathbf{B}^{-1} \mathbf{s} \boldsymbol{1}^H  \mathbf{B}^{-1}}{1- \mathbf{s}^H \mathbf{B}^{-1} \boldsymbol{1}} \Bigr)\boldsymbol{\sigma}^2 \\
& =   \frac{\boldsymbol{1}^H \mathbf{B}^{-1}\boldsymbol{\sigma}^2}{1- \mathbf{s}^H \mathbf{B}^{-1} \boldsymbol{1}}.
    \end{align}
\end{subequations}
Now we can write the power constraint as
$$ \boldsymbol{1}^H \mathbf{B}^{-1} (\boldsymbol{\sigma}^2 - P_t \mathbf{s}) \leq P_t.$$
If we define the vector $\mathbf{b}$ as the vector of the diagonal elements of $\mathbf{B}^{-1}$, then the $k$th element would be
$\mathbf{b}_k=  \gamma/ (| \mathbf{h}_{e_k}^H {\mathbf{u}}_k |^2 + a \sigma_{e_k}^2 \gamma).$
The power constraint is then
\begin{equation}\label{maxmin_cond}
   \sum_k \gamma (\sigma_k^2 + P_t a \sigma_{e_k}^2) / (| \mathbf{h}_{e_k}^H {\mathbf{u}}_k |^2 + a \sigma_{e_k}^2 \gamma) \leq P_t.
\end{equation}
Accordingly, instead of solving the convex problem in \eqref{heurisric} iteratively, we can find the optimal solution by bisection on the SINR target $\gamma$ and checking whether it satisfies \eqref{maxmin_cond} or not.

\subsection{Choice of $a$}\label{sect_a_est}

We stated that for a given $a$, we can solve \eqref{heurisric}. The value of $a$ sets the resulting SINR target(s), and outage probabilities. To obtain the highest goodput, we should solve for all the possible values of $a$, and choose that corresponding to the highest goodput. Since the goodput function increases to a maximum then decreases, then we can use bisection search for the best value of $a$. In each iteration, for the current value of $a$, the outage can be approximated as $e^{-a}$, and the corresponding $\gamma$ can be obtained using bisection to obtain the maximum value that satisfies \eqref{maxmin_cond}. While bisection search is typically fast, having nested bisection search can be of moderate complexity, and we may rather operate using a fixed $a$.

\section{Simulation Results}

For the simulation setup, we will we consider a downlink system in which a BS with $N_t=8$ antennas serves $K=3$ single-antenna users randomly distributed around the BS within a radius of 1km. The large scale fading is described by a path-loss exponent of 3.52 and log-normal shadow fading with 8dB standard deviation, and the small scale fading is modeled using the standard i.i.d. Rayleigh model. The channel estimation error is zero-mean and Gaussian with covariance $\sigma_{e_k}^2 \mathbf{I}$, where $\sigma_{e_k}^2$ is of power -100dBm. The user noise level is -90dBm, and $P_t=40$ watts is the total power constraint. The BS uses ZF directions and the power loading is designed to maximize the minimum rate. In our simulations, we use $d=6$ for the PDF approximation.

\begin{figure}
\begin{center}
    \epsfysize= 2.0in
     \epsffile{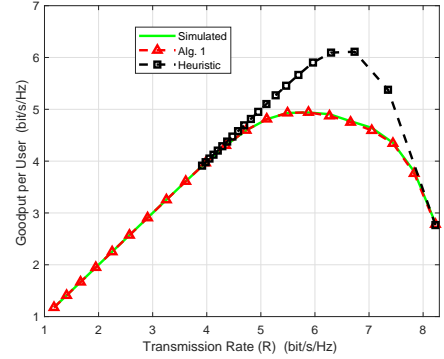}
\caption{The average goodput per user versus $R$ (Single-cell).}\label{fig1}
\end{center}
\end{figure}

In Fig.~\ref{fig1}, we plot the average goodput as defined in \eqref{goodput_obj} versus the transmission rate $R$ for a random set of channel vectors when $\eta=0.3$. We plot the curves for the theoretical goodput values as obtained from Alg.~\ref{Alg1}, and once again using Monte Carlo simulations by producing 2000 random error vector for each SINR target and computing the corresponding goodput. We observe that the PDF approximation using Alg.~\ref{Alg1} provides an excellent approximation for the outage probability, and, accordingly, the goodput all over the transmission rate range.  Note that, for this set of channels, the expected max-min rate value when no uncertainty is present is 8.23 bits/s/Hz while the actual goodput at the user side for that rate is 2.77 bits/s/Hz. We observe that there is significant gap between what is promised by the max-min rate algorithm compared to what is actually delivered to the user side. We also plot the heuristic robust approach for different values of the scale factor $a$ ranging from one to twenty.

\begin{figure}
\begin{center}
    \epsfysize= 2.0in
     \epsffile{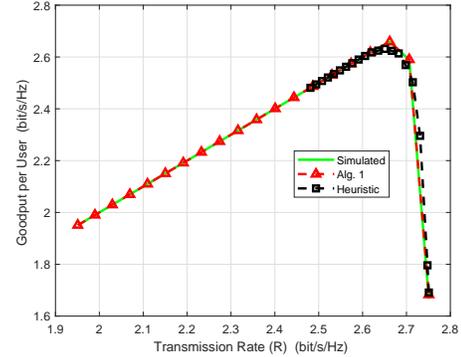}
\caption{The average goodput per user versus $R$ (Multi-cell).}\label{fig2}
\end{center}
\end{figure}

In Fig.~\ref{fig2}, we add the inter-cell interference from the six surrounding BSs to the users' noise level. The other BSs have the same parameters and their centers are 2km away from the center BS. While the significantly higher interference makes the rates lower, we can still observe the same typical performance as that explained for Fig.~\ref{fig1}. In addition, we can see that since the noise is far more dominant now, combating the relatively smaller channel error costs a smaller decrease in the rate. In other words, we only need to slightly decrease the transmission rate  to obtain the maximum goodput.

\begin{figure}
\begin{center}
    \epsfysize= 2.0in
     \epsffile{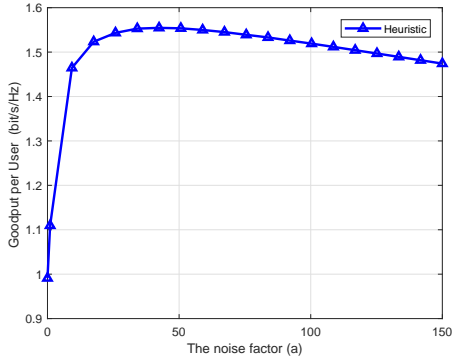}
\caption{The average goodput per user versus the scale $a$.}\label{fig3}
\end{center}
\end{figure}

In Fig.~\ref{fig3}, we show the goodput performance of the heuristic for different values of $a$ for the multi-cell case. We observe that while the delivered rates can be low for small $a$, the performance is almost flat for a range of $a$. Accordingly, obtaining a good performance for a fixed $a$ is possible.

In Fig.~\ref{fig1} (or ~\ref{fig2}), we showed the typical behavior for a set of channels. Now in Table \ref{table1}, we show the average goodput per user for 100 set of channels for each case. For each set, the highest goodput is selected for the theoretical Alg.~\ref{Alg1}, and the heuristic approach in Sect.~\ref{sect_heur_maxmin}. We also show the performance of the heuristic algorithm for $a$ that provides the maximum goodput, and for $a=1 $ which is the original heuristic. We also show the performance when $a$ is estimated as shown in Sect.~\ref{sect_a_est}. We observe that Sect.~\ref{sect_a_est} provides better results when the noise is low, as the outage estimation relies on the interference being dominant.  We can see that, for each case, there is a fixed factor $a$ that provides performance quite close to the best case. We also provide the actual goodput for the max-min rate algorithm that does not take into account the channel uncertainty, versus its claimed performance. Again, this marks a great gap, and suggests that the goodput metric can provide a more accurate representation for the system performance than the rate.

\begin {table}
\begin{center}
\caption{Average goodput in bits/sec/Hertz versus the noise level}
\begin{tabular}{|c|c|c|c|c|c|c|}
  \hline
  cell & Alg.~\ref{Alg1} &  \ref{sect_heur_maxmin} &  Best $a$ & $a=1$ & \ref{sect_a_est}   & Max. min \\
  \hline
   Single & 7.84 &  8.38 &  8.36 ($a=3$) &  7.15 & 8.43 & 3.42 (11.3)  \\
   \hline
  Multi &  1.6  &  1.58 &  1.55 ($a=42$) &  1.1 &  1.4  & 1 (1.68)  \\
  \hline
\end{tabular}
\label{table1}
\end{center}
\end {table}

\section{Conclusion}

In this paper, we proposed using mathematical tools capable of approximating the PDF of a quadratic form with positive definite matrix to obtain the outage probability for a given set of beamformers and transmission rates. The selection of the transmission rate affects the outage probability and the resulting goodput. We proposed a two-layer approach that can provide significant gains with low-computational cost. The approach is based on controlling the outage using a modification to a well-known heuristic, and obtaining the best goodput for that outage. We showed that there can be significant gaps between the rates promised assuming no channel uncertainty and the actual goodput. In summary, the goodput provides a better performance metric to assess the system performance than the rates.

\end{document}